\documentclass[aps, pra, singlecolumn]{revtex4}
\usepackage{amsmath}
\usepackage{dsfont}
\usepackage{graphicx}
\usepackage{graphicx,amsmath,amssymb}

\newcommand{\abs}[1]{\lvert#1\rvert}
\newcommand{\up}{\uparrow}
\newcommand{\down}{\downarrow}
\newcommand{\uu}{\mathbf{u}}
\newcommand{\rr}{\mathbf{r}}

\newcommand{\mm}{\overline m}
\newcommand{\nn}{\overline n}
\newcommand{\pp}{\overline p}
\newcommand{\kk}{\mathbf{k}}
\newcommand{\qq}{\mathbf{q}}
\newcommand{\ii}{\mathrm{i}}

\begin{document}

\title{ The absence of intraband scattering in a consistent theory of Gilbert damping in metallic ferromagnets}

\author{ D M Edwards}

\address{Department of Mathematics, Imperial College London, London SW7~2BZ, United Kingdom}

%\ead{d.edwards@imperial.ac.uk}

\begin{abstract}
Damping of magnetization dynamics in a ferromagnetic metal, arising from spin-orbit coupling, is usually characterised by the Gilbert parameter $\alpha$. Recent calculations of this quantity, using a formula due to Kambersky, find that it is infinite for a perfect crystal owing to an intraband scattering term which is of third order in the spin-orbit parameter $\xi$. This surprising result conflicts with recent work by Costa and Muniz who study damping numerically by direct calculation of the dynamical transverse susceptibility in the presence of spin-orbit coupling. We resolve this inconsistency by following the approach of Costa and Muniz for a slightly simplified model where it is possible to calculate $\alpha$ analytically. We show that to second order in $\xi$ one retrieves the Kambersky result for $\alpha$, but to higher order one does not obtain any divergent intraband terms. The present work goes beyond that of Costa and Muniz by pointing out the necessity of including the effect of long-range Coulomb interaction in calculating damping for large $\xi$. A direct derivation of the Kambersky formula is given which shows clearly the restriction of its validity to second order in $\xi$ so that no intraband scattering terms appear. This restriction has an important effect on the damping over a substantial range of impurity content and temperature. The experimental situation is discussed.    
 
\end{abstract}

\maketitle

\section{Introduction}\label{sec:intro}

Magnetization dynamics in a ferromagnetic metal is central to the field of spintronics with its many applications. Damping is an essential feature of
magnetization dynamics and is usually treated phenomenologically by means of a Gilbert term in the Landau-Lifshitz-Gilbert equation~\cite{LLP80, G55}. For a system with spin-rotational invariance the uniform precession mode of the magnetization in a uniform external magnetic field is undamped and the fundamental origin of damping in ferromagnetic resonance is spin-orbit coupling (SOC). Early investigations of the effect include those of Kambersky~\cite{K70,  K76, K07} and Korenman and Prange~\cite{KP72}. Kambersky's~\cite{K76}  torque-correlation formula for the Gilbert damping parameter $\alpha$ has been used by several groups~\cite{GIS07, GIS08, GMone09, GMtwo09, L09, UMS11, S12, BCEU14}, some of whom have given alternative derivations. However the restricted validity of this formula, as discussed below, has not been stressed. In this torque-correlation model contributions to $\alpha$ of both intraband and interband electronic transitions are usually considered. The theory is basically developed for a pure metal with the effect of defects and/or phonons introduced as phenomenological broadening of the one-electron states. Assuming that the electron scattering-rate increases with temperature T due to electron-phonon scattering the intraband and interband transitions are found to play a dominant role in low and high T regimes, respectively. The intraband(interband) term is predicted to decrease(increase) with increasing T and to be proportional to $\xi^3$($\xi^2$) where $\xi$ is the SOC parameter. Accordingly $\alpha$ is expected to achieve a minimum at an intermediate T. This is seen experimentally in Ni and hcp Co~\cite{BL74} but not in Fe~\cite{BL74} and FePt~\cite{M15}. The $\xi^2$ dependence of $\alpha$ is well-established at high T ~\cite{S07,H13} but there seems to be no experimental observation of the predicted $\xi^3$ behaviour at low T. The interband $\xi^2$ term in Kambersky's theory can be given a very simple interpretation in terms of second-order perturbation theory~\cite{K07}. A quite different phenomenological approach, not applicable in some unspecified low scattering-rate regime, has been adopted to try and find a physical interpretation of the intraband term~\cite{K07, GIS08}. No acceptable theoretical treatment of damping in this low scattering regime is available because the intraband term of Kambersky's theory diverges to infinity in the zero-scattering limit of a pure metal with translational symmetry at T=0~\cite{GMone09, L09, S12}. We consider it essential to understand the pure metal limit before introducing impurity and phonon scattering in a proper way.

 Costa and Muniz~\cite{CM15} recently studied damping numerically in this limit by direct calculation of the dynamical spin susceptibility in the presence of SOC within the random phase approximation (RPA). They determine $\alpha$ from the linewidth of the uniform (wave-vector $\qq=0$) spin-wave mode which appears as a resonance in the transverse susceptibility. One of the main objects of this paper is to establish some degree of consistency between the work of Kambersky and that of Costa and Muniz. We follow the approach of the latter authors for a slightly simplified model where it is possible to calculate $\alpha$ analytically. We show that to second order in $\xi$ one retrieves the Kambersky result, but to higher order no intraband terms occur, which removes the problem of divergent $\alpha$. To confirm this point, in Appendix~\ref{Ka} we derive the Kambersky formula directly in a way that makes clear its restriction to second order in $\xi$ to which order the divergent terms in $\alpha$ arising from intraband transitions do not appear. This throws open the interpretation of the minimum observed in the temperature dependence of $\alpha$ for Ni and Co.

At this point we may mention an alternative theoretical approach to the calculation of Gilbert damping using scattering theory~\cite{BTB08, SK10}. Starikov et al~\cite{SK10} find that, for  $\mathrm{Ni_{1-x} Fe_x}$ alloys at T=0,  $\alpha$ becomes large near the pure metal limits x=0,1. They attribute this to the Kambersky intraband contribution although no formal correspondence is made between the two approaches.  

   The work of Costa and Muniz~\cite{CM15} follows an earlier paper~\cite{CMLKM10} where it is shown that SOC has the effect of coupling the transverse spin susceptibility to the longitudinal spin susceptibility and the charge response. It is known that a proper calculation of these last two quantities in a ferromagnet must take account of long-range Coulomb interactions~\cite{RBR67, L69, KPS69, KSP73, R78}. The essential role of these interactions is to ensure conservation of particle number. Costa et al~\cite{CM15, CMLKM10} do not consider such interactions but we show here that this neglect is not serious for calculating $\alpha$ with sufficiently small SOC. However in the wider framework of this paper, where mixed charge-spin response is also readily studied, long-range interactions are expected to sometimes play a role. They also come into play, even to second order in $\xi$, when inversion symmetry is broken.
 
  In section~\ref{SDFT} we establish the structure of spin-density response theory in the presence of SOC by means of exact spin-density functional theory in the static limit~\cite{WvB83}. In section~\ref{dynamical susceptibilities} we introduce a spatial Fourier transform and an approximation to the dynamical response is obtained by introducing the frequency dependence of the non-interacting susceptibilities. The theory then has the same structure as in the RPA. Section ~\ref{transversechi} is devoted to obtaining an explicit expression for the transverse susceptibility in terms of the non-interacting susceptibilities. Expressions for these, in the presence of SOC, are obtained within the tight-binding approximation in section~\ref{noninteracting chi}. In section~\ref{linewidth} we consider the damping of the resonance in the q=0 transverse susceptibility and show how the present approach leads to the Kambersky formula for the Gilbert damping parameter $\alpha$ where this is valid, namely to second order in the SOC parameter $\xi$. We do not give an explicit formula for $\alpha$ beyond this order but it is clear that no intraband terms appear. In section~\ref{expt} some experimental aspects are discussed with suggestions for future work. The main conclusions are summarized in section~\ref{discussion}. 

\section{Spin-density functional theory with spin-orbit coupling}\label{SDFT}
The Kohn-Sham equation takes the form
\begin{equation}\label{Kohn-Sham}
\sum_{\sigma^{\prime}}[-\delta_{\sigma\sigma^{\prime}}(\hbar^2/2m)\nabla^2+V_{\sigma\sigma^{\prime}}^{eff}(\rr)
+H_{\sigma\sigma^{\prime}}^{so}]\phi_{n\sigma^{\prime}}(\rr)=\epsilon_{n}\phi_{n\sigma}(\rr)
\end{equation}
with the spin index $\sigma=\uparrow,\downarrow$ corresponding to quantization along the direction of the ground-state magnetization in a ferromagnet. This may be written in $2\times2$ matrix form with eigenvectors $(\phi_{n\uparrow},\phi_{n\downarrow})^T$. The density matrix is defined in terms of the spin components $\phi_{n\sigma}(\rr)$ of the one-electron orbitals by
\begin{equation}\label{density matrix}
n_{\sigma\sigma^{\prime}}=\sum_{n}\phi_{n\sigma}(\rr)\phi_{n\sigma^{\prime}}(\rr)^{*}\theta(\mu_0-\epsilon_n)
\end{equation}
where $\theta(x)$ is the unit step function and $\mu_0$ is the chemical potential. The electron density is given by
\begin{equation}\label{density}
\rho(\rr)=\sum_{\sigma}n_{\sigma\sigma}(\rr)=\sum_{n\sigma}\abs{\phi_{n\sigma}(\rr)}^2\theta(\mu_0-\epsilon_n)
\end{equation}
and the effective potential in~\eqref{Kohn-Sham} is
\begin{equation}\label{Veff}
V_{\sigma\sigma^{\prime}}^{eff}(\rr)=w_{\sigma\sigma^{\prime}}(\rr)
+\delta_{\sigma\sigma^{\prime}}\int d^{3} r^{\prime} \rho(\rr^{\prime})v(\rr-\rr^{\prime})+v_{\sigma\sigma^{\prime}}
^{xc}(\rr)
\end{equation}
where $w_{\sigma\sigma^{\prime}}(\rr)$ is the external potential due to the crystal lattice and any magnetic fields and $v(\rr)=e^2/\abs{\rr}$ is the Coulomb potential. The exchange-correlation potential $v_{\sigma\sigma^{\prime}}
^{xc}(\rr)$ is defined as $\delta E_{xc}/\delta n_{\sigma\sigma^{\prime}}(\rr)$, a functional derivative of the exchange-correlation energy $E_{xc}$. The term $H_{\sigma\sigma^{\prime}}^{so}$ in~\eqref{Kohn-Sham} is the SOC energy. A small external perturbation $\delta w_{\sigma\sigma^{\prime}}$, for example due to a magnetic field, changes the effective potential to $V^{eff}+\delta V^{eff}$, giving rise to new orbitals and hence to a change in density matrix $\delta n_{\sigma\sigma^{\prime}}$. The equation
\begin{equation}\label{deltan}
\delta  n_{\sigma\sigma^{\prime}}(\rr)=-\Omega^{-1}\sum_{\sigma_{1}\sigma_{1}^{\prime}}
\int d^3 r_1 \chi_{\sigma\sigma^{\prime}\sigma_{1}\sigma_{1}^{\prime}}^{0}(\rr,\rr_{1})\delta
V_{\sigma_{1}\sigma_{1}^{\prime}}^{eff}(\rr_{1}),
 \end{equation}
 where $\Omega$ is the volume of the sample, defines a non-interacting response function $\chi^{0}$ and the full response function $\chi$ is defined by
\begin{equation}\label{chi defined}
\delta  n_{\sigma\sigma^{\prime}}(\rr)=-\Omega^{-1}\sum_{\tau\tau^{\prime}}
\int d^3 r^{\prime} \chi_{\sigma\sigma^{\prime}\tau\tau^{\prime}}(\rr,\rr^{\prime})\delta
w_{\tau\tau^{\prime}}(\rr^{\prime}).
 \end{equation}
 From ~\eqref{Veff}
 \begin{equation}\label{deltaVeff}
 \delta V_{\sigma_{1}\sigma_{1}^{\prime}}^{eff}(\rr_{1})=\delta w_{\sigma_{1}\sigma_{1}^{\prime}}(\rr_{1})
+\sum_{\sigma_{2}\sigma_{2}^{\prime}}\int d^3 r_{2}[v(\rr_{1}-\rr_{2})\delta_{\sigma_{1}\sigma_{1}^{\prime}}
\delta_{\sigma_{2}\sigma_{2}^{\prime}}+
\frac{ \delta v_{\sigma_{1}\sigma_{1}^{\prime}}^{xc}(\rr_{1})}{\delta n_{\sigma_{2}\sigma_{2}^{\prime}}(\rr_{2})}]
\delta n_{\sigma_{2}\sigma_{2}^{\prime}}(\rr_{2})
\end{equation}
and we may write
\begin{equation}\label{exchangeK}
\frac{ \delta v_{\sigma_{1}\sigma_{1}^{\prime}}^{xc}(\rr_{1})}{\delta n_{\sigma_{2}\sigma_{2}^{\prime}}(\rr_{2})}
=\frac{\delta^{2} E_{xc}}{\delta n_{\sigma_{2}\sigma_{2}^{\prime}}(\rr_{2})\delta n_{\sigma_{1}\sigma_{1}^{\prime}}(\rr_{1})}=K_{\sigma_{1}\sigma_{1}^{\prime}\sigma_{2}\sigma_{2}^{\prime}}(\rr_{1},
\rr_{2}).
\end{equation} 
Combining ~\eqref{deltan} -~\eqref{exchangeK} we find the following integral equation for the spin-density response function $ \chi_{\sigma\sigma^{\prime}\tau\tau^{\prime}}(\rr,\rr^{\prime})$:
\begin{equation}\label{chi integral eqn}
\begin{split}
\chi_{\sigma\sigma^{\prime}\tau\tau^{\prime}}(\rr,\rr^{\prime})=
  \chi_{\sigma\sigma^{\prime}\tau\tau^{\prime}}^{0}(\rr,\rr^{\prime})-
  (\Omega)^{-1}\sum_{\sigma_{1}\sigma_{1}^{\prime}}\sum_{\sigma_{2}\sigma_{2}^{\prime}}\int d^3 r_{1}\int d^{3} r_{2}\: \chi_{\sigma\sigma^{\prime}\sigma_{1}\sigma_{1}^{\prime}}^{0}(\rr,\rr_{1})
[v(\rr_{1}-\rr_{2})\delta_{\sigma_{1}\sigma_{1}^{\prime}}\delta_{\sigma_{2}\sigma_{2}^{\prime}}
+K_{\sigma_{1}\sigma_{1}^{\prime}\sigma_{2}\sigma_{2}^{\prime}}(\rr_{1},\rr_{2})]\\
 \chi_{\sigma_{2}\sigma_{2}^{\prime}\tau\tau^{\prime}}(\rr_{2},\rr^{\prime}).
 \end{split}
 \end{equation}
 This equation is a slight generalisation of that given by Williams and von Barth ~\cite{WvB83}. In the static limit it is formally exact although the exchange-correlation energy $E_{xc}$ is of course not known exactly. In the next section we generalise the equation to the dynamical case approximately by introducing the frequency dependence of the non-interacting response functions $\chi^0$, and also take a spatial Fourier transform. In the case where SOC is absent the result is directly compared with results obtained using the RPA.
\section{Dynamical susceptibilities in the presence of spin-orbit coupling and long-range Coulomb interaction}\label{dynamical susceptibilities}

In general the response functions $\chi(\rr,\rr^{\prime})$ are not functions of $\rr-\rr^{\prime}$ and a Fourier representation of~\eqref{chi integral eqn} for a spatially periodic system involves an infinite number of reciprocal lattice vectors. There are two cases where this complication is avoided. The first is a homogeneous electron gas and the second is in a tight-binding approximation with a restricted atomic basis. We may then introduce Fourier transforms of the form $\chi(\rr)=\sum_{\qq}\chi(\qq)e^{\ii\qq\cdot\rr}$ or $\chi(\qq)=(\Omega)^{-1}\int d^{3}r\chi(\rr)e^{-\ii\qq\cdot\rr}$ and write \eqref{chi integral eqn} as
\begin{equation}\label{chiq}
\chi_{\sigma\sigma^{\prime}\tau\tau^{\prime}}(\qq,\omega)=
  \chi_{\sigma\sigma^{\prime}\tau\tau^{\prime}}^{0}(\qq,\omega)-
  \sum_{\sigma_{1}\sigma_{1}^{\prime}}\sum_{\sigma_{2}\sigma_{2}^{\prime}}  \chi_{\sigma\sigma^{\prime}\sigma_{1}\sigma_{1}^{\prime}}^{0}(\qq,\omega)
V_{\sigma_{1}\sigma_{1}^{\prime}\sigma_{2}\sigma_{2}^{\prime}}(\qq)
 \chi_{\sigma_{2}\sigma_{2}^{\prime}\tau\tau^{\prime}}(\qq,\omega),
 \end{equation}
 where we have also introduced the $\omega$ dependence of $\chi$ as indicated at the end of the last section. Here $V(\qq)$ is an ordinary Fourier transform, without a factor $(\Omega)^{-1}$, so that
\begin{equation}\label{Vq}
V_{\sigma_{1}\sigma_{1}^{\prime}\sigma_{2}\sigma_{2}^{\prime}}(\qq)
=v(\qq)\delta_{\sigma_{1}\sigma_{1}^{\prime}}\delta_{\sigma_{2}\sigma_{2}^{\prime}}
+K_{\sigma_{1}\sigma_{1}^{\prime}\sigma_{2}\sigma_{2}^{\prime}},
\end{equation}
where $v(\qq)=4\pi e^{2}/q^{2}$ is the usual Fourier transform of the Coulomb interaction and the second term is independent of $\qq$ since $K$ is a short-range spatial interaction. In the gas case it is proportional to a delta-function $\delta(\rr-\rr^{\prime})$ in the local-density approximation (LDA) ~\cite{WvB83} and in tight-binding it can be taken as an on-site interaction. In both cases $K$ may be expressed in terms of a parameter $U$ as
\begin{equation}\label{K}
K_{\sigma_{1}\sigma_{1}^{\prime}\sigma_{2}\sigma_{2}^{\prime}}
=-U[\delta_{\sigma_{1}\sigma_{1}^{\prime}}\delta_{\sigma_{2}\sigma_{2}^{\prime}}\delta_{\sigma_{1}\sigma_{2}}
+\delta_{\overline{\sigma}_{1}\sigma_{1}^{\prime}}\delta_{\overline{\sigma}_{2}\sigma_{2}^{\prime}}
\delta_{\sigma_{1}^{\prime}\sigma_{2}}]
\end{equation}
where $\overline{\sigma}=\downarrow,\uparrow$ for $\sigma=\uparrow,\downarrow$. in the tight-binding case this form of $K$ corresponds to a simple form of interaction which leads to a rigid exchange splitting of the bands (~\cite{E84},~\cite{CMLKM10}). This is only appropriate for transition metals in a model with d bands only, hybridization with s and p bands being neglected. We adopt this model in order to obtain transparent analytic results as far as possible. Although not as realistic as "first-principles" models of the electronic structure it has been used, even with some quantitative success, in treating the related problem of magnetocrystalline anisotropy in Co/Pd structures as well as pure metals~\cite{CE97}. In~\eqref{chiq} the response functions $\chi$ are per unit volume in the gas case but, more conveniently, may be taken as per atom in the tight-binding case with $v(\qq)$ modified to $v(\qq)=4\pi e^{2}/(q^{2}\Omega_a)$ where $\Omega_a$ is the volume per atom.

To show how equations~\eqref{chiq} -~\eqref{K} are related to RPA we examine two examples in the absence of SOC. First consider the transverse susceptibility $\chi_{\down\up\up\down}(\qq,\omega)$ which is more usually denoted by $\chi_{-+}(\qq,\omega)$.
Equation~\eqref{chiq} becomes
\begin{equation}\label{chitransverse}
\chi_{\down\up\up\down}=\chi_{\down\up\up\down}^{0}
-\chi_{\down\up\up\down}^{0} V_{\up\down\down\up}\chi_{\down\up\up\down}
\end{equation}
and, from~\eqref{Vq} and~\eqref{K}, $V_{\up\down\down\up}= K_{\up\down\down\up}=-U$. Hence
\begin{equation}\label{IKK}
\chi_{\down\up\up\down}=\chi_{\down\up\up\down}^{0}(1-U\chi_{\down\up\up\down}^{0})^{-1}
\end{equation}
which is just the RPA result of Izuyama et al~\cite{IKK63} for a single-orbital Hubbard model and of Lowde and Windsor~\cite{LW70} for a five-orbital d-band model. Clearly in the absence of SOC the Coulomb interaction $v(\qq)$ plays no part in the transverse susceptibility, as is well-known. A more interesting case is the longitudinal susceptibility denoted by $\chi_{mm}$ in the work of Kim et al (~\cite{KPS69},~\cite{KSP73}) and in~\cite{WvB83}. This involves only the response functions $\chi_{\sigma\sigma\tau\tau}$ which we abbreviate to $\chi_{\sigma\tau}$. In fact~\cite{WvB83} 
\begin{equation}\label{chimm1}
\chi_{mm}=\chi_{\up\up}+\chi_{\down\down}-\chi_{\up\down}-\chi_{\down\up}.
\end{equation}
Without SOC $\chi_{\sigma\tau}^{0}$ takes the form $\chi_{\sigma}^{0}\delta_{\sigma\tau}$ and ~\eqref{chiq} becomes
\begin{equation}\label{chisigmatau}
\chi_{\sigma\tau}=\chi_{\sigma}^{0}\delta_{\sigma\tau}
-\sum_{\sigma_{2}}\chi_{\sigma}^{0} V_{\sigma\sigma_{2}}\chi_{\sigma_{2}\tau}
\end{equation}
with $ V_{\sigma\sigma_{2}}=v(\qq)-U\delta_{\sigma\sigma_2}$. On solving the $2\times2$ matrix equation~\eqref{chisigmatau} for $\chi_{\sigma\tau}$, and using~\eqref{chimm1}, we find the longitudinal susceptibility in the form
\begin{equation}\label{chimm2}
\chi_{mm}=\frac{\chi_{\up}^{0}+\chi_{\down}^{0}-2[U-2v(\qq)]\chi_{\up}^{0}\chi_{\down}^{0}}
{1+(\chi_{\up}^{0}+\chi_{\down}^{0})[v(\qq)-U]+U[U-2v(\qq)]\chi_{\up}^{0}\chi_{\down}^{0}}
\end{equation}
 which agrees with the RPA result that Kim et al (~\cite{KPS69},~\cite{KSP73})found for a single-orbital model. The Coulomb interaction $v(\qq)$ is clearly important, particularly for the uniform susceptibility with q=0, where $v\rightarrow\infty$. It plays an essential role in enforcing particle conservation and hence in obtaining the correct result of Stoner theory. In view of the correspondence between our approach and the RPA method it seems likely that when SOC is included our procedure using equations~\eqref{chiq} -~\eqref{K} should be almost equivalent to that of Costa and Muniz~\cite{CM15} in the case of a model with d-bands only. However our inclusion of the long-range Coulomb interaction will modify the results.
\section{An explicit expression for the transverse susceptibility}\label{transversechi}
In this section we obtain an explicit expression for the transverse susceptibility $\chi_{\down\up\up\down}$ in terms of the non-interacting response functions $\chi^{0}$. We consider equation~\eqref{chiq} as an equation between $4\times4$ matrices where $\sigma\sigma^{\prime}=\down\up,\up\down,\up\up,\down\down$ labels the rows in that order and $\tau\tau^{\prime}$ labels columns similarly. The formal solution of~\eqref{chiq} is then
\begin{equation}\label{chimatrix}
\chi=(1+\chi^{0}V)^{-1}\chi^{0}.
\end{equation}
This expression could be used directly as the basis of a numerical investigation similar to that of Costa and Muniz. However we wish to show that the present approach leads to a Gilbert damping parameter $\alpha$ in agreement with the Kambersky formula, to second order in the SOC parameter $\xi$ where Kambersky's result is valid. This requires some quite considerable analytic development of~\eqref{chimatrix}.

First we partition each matrix in~\eqref{chimatrix} into four $2\times2$ matrices. Thus from~\eqref{Vq} and~\eqref{K}
\begin{equation}\label{Vmatrix}
V=\begin{pmatrix}V_1 & 0 \\
0 & V_2 \\
\end{pmatrix}
\end{equation}
with
\begin{equation}\label{V1V2}
V_1=\begin{pmatrix}0 & -U \\
-U & 0 \\
\end{pmatrix},\thickspace V_2=\begin{pmatrix}v-U & v \\
v & v-U \\
\end{pmatrix}.
\end{equation}
Also
\begin{equation}\label{chimatrixpartitioned}
\chi=\begin{pmatrix}\chi_{11} & \chi_{12}\\
\chi_{21} & \chi_{22}
\end{pmatrix}
\end{equation}
and similarly for $\chi^{0}$. If we write
\begin{equation}\label{ABCD}
1+\chi^{0} V=\begin{pmatrix}1+\chi_{11}^{0}V_1 & \chi_{12}^{0}V_2\\
\chi_{21}^{0}V_1 & 1+\chi_{22}^{0}V_2 \end{pmatrix}=
\begin{pmatrix}A & B\\C & D 
\end{pmatrix}
\end{equation}
~\eqref{chimatrix} becomes
\begin{equation}\label{chimatrixexplicit}
\chi=\begin{pmatrix}S^{-1} & -S^{-1}BD^{-1}\\
-D^{-1}CS^{-1} & \quad D^{-1}+D^{-1}CS^{-1}BD^{-1}
\end{pmatrix}\begin{pmatrix}\chi_{11}^{0} & \chi_{12}^{0}\\
\chi_{21}^{0} & \chi_{22}^{0}
\end{pmatrix}
\end{equation}
where
\begin{equation}\label{S1}
S=A+BD^{-1}C.
\end{equation}
The transverse susceptibility $\chi_{\down\up\up\down}$ in which we are interested is the top right-hand element of $\chi_{11}$ so this is the quantity we wish to calculate. From~\eqref{chimatrixexplicit} 
\begin{equation}\label{chi11}
\chi_{11}=S^{-1}(\chi_{11}^{0}-BD^{-1}\chi_{21}^{0})
\end{equation}
and, from~\eqref{S1} and~\eqref{ABCD},
\begin{equation}\label{S2}
S=1+\chi_{11}^{0}V_1-\chi_{12}^{0}(V_2^{-1}+\chi_{22}^{0})^{-1}\chi_{21}^{0}V_1.
\end{equation}
The elements of the $2\times2$ matrix S are calculated by straight-forward algebra and 
\begin{equation}\label{S11}
\begin{split}
S_{11}=1-U\chi_{\down\up\up\down}^{0}
+(U/\Lambda)[(X+\chi_{\down\down\down\down}^{0})\chi_{\up\up\up\down}^{0}\chi_{\down\up\up\up}^{0}
-(Y+\chi_{\up\up\down\down}^{0})\chi_{\down\down\up\down}^{0}\chi_{\down \up \up\up}^{0}\\-(Y+\chi_{\down\down\up\up}^{0})\chi_{\up\up\up\down}^{0}\chi_{\down\up\down\down}^{0}
+(X+\chi_{\up\up\up\up}^{0})\chi_{\down\down\up\up}^{0}\chi_{\down\up\down\down}^{0}]
\end{split}
\end{equation}
where
\begin{equation}\label{XY}
X=[v-U]/[U(U-2v)],\quad Y=-v/[U(U-2v)]
\end{equation}
and 
\begin{equation}\label{Lamda}
\Lambda=(X+\chi_{\up\up\up\up}^{0})(X+\chi_{\down\down\down\down}^{0})
-(Y+\chi_{\up\up\down\down}^{0})(Y+\chi_{\down\down\up\up}^{0})
\end{equation}
The other three elements of $S$ are given in Appendix~\ref{Selements}. The transverse susceptibility is obtained from~\eqref{chi11} as
\begin{equation}\label{chi-+}
\chi_{\down\up\up\down}=[S_{22}(\chi_{11}^{0}-BD^{-1}\chi_{21}^{0})_{12}
-S_{12}(\chi_{11}^{0}-BD^{-1}\chi_{21}^{0})_{22}]/(S_{11}S_{22}-S_{12}S_{21})
\end{equation}
and
\begin{equation}\label{BD}
BD^{-1}=\chi_{12}^{0}(V_2^{-1}+\chi_{22}^{0})^{-1}.
\end{equation}
A comparison of the fairly complex equation above for the transverse susceptibility with the simple well-known result ~\eqref{IKK} shows the extent of the new physics introduced by SOC. This is due to the coupling of the transverse susceptibility to the longitudinal susceptibility and the charge response, both of which involve the long-range Coulomb interaction.

To proceed further it is necessary to specify the non-interacting response functions $\chi_{\sigma\sigma^{\prime}\sigma_{1}\sigma_{1}^{\prime}}^{0}$ which occur throughout the equations above.  
\section{The non-interacting response functions}\label{noninteracting chi}

In the tight-binding approximation the one-electron basis functions are the Bloch functions
\begin{equation}\label{Bloch}
\vert \kk\mu\sigma\rangle=N^{-1/2}\sum_j  {\rm e}^{\ii\kk\cdot\mathbf{R}_j}\vert j\mu\sigma\rangle
\end{equation}
where $j$ and $\mu$ are the site and orbital indices, respectively, and $N$ is the number of atoms in the crystal. The Hamiltonian in the Kohn-Sham equation now takes the form
\begin{equation}\label{Heff}
H^{eff}=\sum_{\kk\mu\nu\sigma}(T_{\mu\nu}(\kk)
+V_{\sigma}^{eff}\delta_{\mu\nu})c_{\kk\mu\sigma}^{\dagger}c_{\kk\nu\sigma}+H^{so}
\end{equation} 
where $T_{\mu\nu}$ corresponds to electron hopping and 
\begin{equation}\label{Vsigmaeff}
V_{\sigma}^{eff}=-(\sigma/2)(\Delta+b_{ex})
\end{equation}
where $\sigma=1,-1$ for spin $\up,\down$ respectively. Here $\Delta=2U\langle S^{z}\rangle/N$ where $S^{z}$ is the total spin angular momentum, in units of $\hbar$, and the Zeeman splitting $ b_{ex}=2\mu_{B}B_{ex}$, where $B_{ex}$ is the external magnetic field and $\mu_B$ is the Bohr magneton. The spin-orbit term $H^{so}=\xi\sum_{j} \mathbf{L}_{j}\cdot\mathbf{S}_j$ takes the second-quantized form
\begin{equation}\label{Hso}
H^{so}=(\xi/2)\sum_{\kk\mu\nu}[L_{\mu\nu}^{z}(c_{\kk\mu\up}^{\dagger}
c_{\kk\nu\up}-c_{\kk\mu\down}^{\dagger}c_{\kk\nu\down})+L_{\mu\nu}^{+}c_{\kk\mu\down}^{\dagger}c_{\kk\nu\up}
+L_{\mu\nu}^{-}c_{\kk\mu\up}^{\dagger}c_{\kk\nu\down}]
\end{equation}
where $L_{\mu\nu}^{z}, L_{\mu\nu}^{\pm}$ are matrix elements of the atomic orbital angular momentum operators $L^{z}, L^{\pm}=L^{x}\pm \ii L^{y}$ in units of $\hbar$. Within the basis of states~\eqref{Bloch} eigenstates of $H^{eff}$ take the form
\begin{equation}\label{eigenstates}
\vert \kk n\rangle=\sum_{\mu\sigma}a_{n\mu}^{\sigma}(\kk)\vert \kk\mu\sigma\rangle,
\end{equation}
and satisfy the equation
\begin{equation}\label{eigenstate equation}
H^{eff}\vert\kk n\rangle=E_{\kk n}\vert\kk n\rangle.
\end{equation}
Thus
\begin{equation}\label{cdagger}
c_{\kk\mu\sigma}^{\dagger}=\sum_{n} a_{n\mu}^{\sigma}(\kk)^{*}c_{\kk n}^{\dagger}
\end{equation}
where $c_{\kk n}^{\dagger}$ creates the eigenstate $\vert \kk n\rangle$.

The non-interacting response function $ \chi_{\sigma\sigma^{\prime}\sigma_{1}\sigma_{1}^{\prime}}^{0}(\qq,\omega)$ is conveniently expressed as the Fourier transform of a retarded Green function by the Kubo formula
\begin{equation}\label{chi0}
\chi_{\sigma\sigma^{\prime}\sigma_{1}\sigma_{1}^{\prime}}^{0}(\qq,\omega)
 =\sum_{\kk}\langle\langle\sum_{\mu}c_{\kk+\qq\mu\sigma}^{\dagger}c_{\kk\mu\sigma^{\prime}};
 \sum_{\nu}c_{\kk\nu\sigma_{1}}^{\dagger}c_{\kk+\qq\nu\sigma_{1}^{\prime}}\rangle\rangle_{\omega}^{0}
 \end{equation}
 where the right-hand side is to be evaluated using the one-electron Hamiltonian $H^{eff}$. Consequently, using~\eqref{cdagger}, we have
\begin{equation}\label{chi0explicit}
\begin{split}
 \chi_{\sigma\sigma^{\prime}\sigma_{1}\sigma_{1}^{\prime}}^{0}(\qq,\omega)
 =\sum_{\kk\mu\nu}\sum_{mn}a_{m\mu}^{\sigma}(\kk+\qq)^{*}a_{n\mu}^{\sigma^{\prime}}(\kk)
 a_{n\nu}^{\sigma_{1}}(\kk)^{*}a_{m\nu}^{\sigma_{1}^{\prime}}(\kk+\qq)
  \langle\langle c_{\kk+\qq m}^{\dagger} c_{\kk n};
 c_{\kk n}^{\dagger}c_{\kk+\qq m}\rangle\rangle_{\omega}^{0}\\
 =N^{-1}\sum_{\kk\mu\nu}\sum_{mn}a_{m\mu}^{\sigma}(\kk+\qq)^{*}a_{n\mu}^{\sigma^{\prime}}(\kk)
 a_{n\nu}^{\sigma_{1}}(\kk)^{*}a_{m\nu}^{\sigma_{1}^{\prime}}(\kk+\qq)
 \frac{f_{\kk n}-f_{\kk+\qq m}}{E_{\kk+\qq m}-E_{\kk n}-\hbar\omega+\ii\eta}.
 \end{split}
 \end{equation}
 The last step uses the well-known form of the response function per atom for a non-interacting Fermi system (e.g.~\cite{DS98}) and $\eta$ is a small positive constant which ultimately tends to zero. The occupation number $f_{\kk n}=F(E_{\kk n}-\mu_{0})$ where $F$ is the Fermi function with chemical potential $\mu_{0}$. Clearly for $\qq=0$ the concept of intraband transitions ($m=n$), frequently introduced in discussions of the Kambersky formula, never arises  for finite $\omega$ since the difference of the Fermi functions in the numerator of~\eqref{chi0explicit} is zero. Equation~\eqref{chi0explicit} may be written in the form
\begin{equation}\label{chi02}
\chi_{\sigma\sigma^{\prime}\sigma_{1}\sigma_{1}^{\prime}}^{0}(\qq,\omega)
=N^{-1}\sum_{\kk mn}B_{mn}^{\sigma\sigma^{\prime}}(\kk,\qq)B_{mn}^{{\sigma_1}^{\prime}\sigma_{1}}(\kk,\qq)^{*}
 \frac{f_{\kk n}-f_{\kk+\qq m}}{E_{\kk+\qq m}-E_{\kk n}-\hbar\omega+\ii\eta}
\end{equation}
where
\begin{equation}\label{B}
B_{mn}^{\sigma\sigma^{\prime}}(\kk,\qq)=\sum_{\mu}a_{m\mu}^{\sigma}(\kk+\qq)^{*}a_{n\mu}^{\sigma^{\prime}}(\kk).
\end{equation}
\section{Ferromagnetic resonance linewidth; the Kambersky formula}\label{linewidth}

We now consider the damping of the ferromagnetic resonance in the $q=0$ transverse susceptibility. The present approach, like the closely-related one of Costa and Muniz~\cite{CM15}, is valid for arbitrary strength of the SOC and can be used as the basis of numerical calculations, as performed by the latter authors. However it is important to show analytically that the present method leads to the Kambersky~\cite{K76} formula for the Gilbert damping parameter where this is valid, namely to second order in the SOC parameter $\xi$. This is the subject of this section.

It is useful to consider first the case without SOC ($\xi=0$). The eigenstates $n$ of $H^{eff}$ then have a definite spin and may be labelled $\nn\sigma$. It follows from~\eqref{chi0explicit} that $\chi_{\sigma\sigma^{\prime}\sigma_{1}\sigma_{1}^{\prime}}^{0}
\propto\delta_{\sigma\sigma_{1}^{\prime}}\delta_{\sigma^{\prime}\sigma_{1}}$. Hence $\chi_{12}^{0}=0$ and, from~\eqref{BD}, $BD^{-1}=0$. Also, from Appendix B, $S_{12}=S_{21}=0$. Thus,~\eqref{chi-+} reduces to~\eqref{IKK} as it should. Considering $\chi_{\down\up\up\down}^{0}(0,\omega)$, given by~\eqref{chi0explicit} and~\eqref{chi02}, we note that state $m$ is pure $\down$ spin, labelled by $\mm\down$, and $n$ is pure $\up$, labelled by $\nn\up$. Hence for $\xi=0$ \begin{equation}\label{Bmn} 
B_{mn}^{\down\up}(\kk,0)=\sum_{\mu}\langle\kk\mm\vert\kk\mu\rangle\langle\kk\mu\vert\kk\nn\rangle
=\delta_{\mm\thinspace\nn}
\end{equation}
from closure. Thus
\begin{equation}\label{transversechi0}
\chi_{\down\up\up\down}^{0}(0,\omega)=N^{-1}\sum_{\kk\nn} \frac{f_{\kk \nn\up}-f_{\kk \nn\down}}{E_{\kk \nn\down}-E_{\kk \nn\up}-\hbar\omega+\ii\eta}
\end{equation}
and it follows from~\eqref{Vsigmaeff} that $E_{\kk\nn\sigma}$ may be written as 
\begin{equation}\label{splitband}
E_{\kk\nn\sigma}=E_{\kk\nn}-(\sigma/2)(\Delta+b_{ex}).
\end{equation}
Hence we find from~\eqref{IKK} that for $\xi=0$
\begin{equation}\label{spinwavepole}
\chi_{\down\up\up\down}(0,\omega)=(2\langle S^{z}\rangle/N)(b_{ex}-\hbar\omega+\ii\eta)^{-1}.
\end{equation}
Thus, as $\eta\rightarrow 0$, $\Im\chi_{\down\up\up\down}(0,\omega)$ has a sharp delta-function resonance at $\hbar\omega=b_{ex}$ as expected.

When SOC is included $\hbar\omega$ acquires an imaginary part that corresponds to damping. We now proceed to calculate this imaginary part to $O(\xi^2)$. To do this we can take $\xi=0$ in the numerator of~\eqref{chi-+} so that \begin{equation}\label{chitransverse2}
\chi_{\down\up\up\down}(0,\omega)=\chi_{\down\up\up\down}^{0}(0,\omega)/
(S_{11}-S_{12}S_{21}/S_{22})
\end{equation}
In fact $S_{12}$ and $S_{21}$ are both  $O(\xi^2)$ while $S_{22}$ is $O(1)$. Thus to obtain $\hbar\omega$ to  $O(\xi^{2})$ we need only solve $S_{11}=0$. Furthermore all response functions such as $\chi_{\up\up\up\down}^{0}$, with all but one spins in the same direction, are zero for $\xi=0$ and need only be calculated to $O(\xi)$ in~\eqref{S11}. We show below that to this order they vanish, so that to $O(\xi^2)$ the last term in $S_{11}$ is zero and we only have to solve the equation $1-U\chi_{\down\up\up\down}^{0}=0$ for $\hbar\omega$. This means that to second order in $\xi$ the shift in resonance frequency and the damping do not depend on the long-range Coulomb interaction.

To determine $\chi_{\up\up\up\down}^{0}(0,\omega)$ to first order in $\xi$ from~\eqref{chi0explicit} we notice that states $n$ must be pure $\up$ spin, that is $\vert\kk n\rangle=\vert\kk\nn\up\rangle$, while states $m$ must be calculated using perturbation theory. The latter states may be written
\begin{equation}\label{perturbed state1}
\vert\kk\mm 1\rangle=\vert\kk\mm\up\rangle-\xi\sum_{\pp\sigma}
\frac{\langle\kk\pp\sigma\vert h^{so}\vert\kk\mm\up\rangle}{E_{\kk\mm\up}-E_{\kk\pp\sigma}}
\vert\kk\pp\sigma\rangle
\end{equation}
\begin{equation}\label{perturbed state2}
\vert\kk\mm 2\rangle=\vert\kk\mm\down\rangle-\xi\sum_{\pp\sigma}
\frac{\langle\kk\pp\sigma\vert h^{so}\vert\kk\mm\down\rangle}{E_{\kk\mm\down}-E_{\kk\pp\sigma}}
\vert\kk\pp\sigma\rangle,
\end{equation}
where we have put $H^{so}=\xi h^{so}$, and to first order in $\xi$,
\begin{equation}\label{chi03}
\chi_{\up\up\up\down}^{0}=\frac{1}{N}\sum_{\kk\mu\nu}\sum_{\mm\,\nn}
({a_{\mm 1\mu}^{\up}}^{*}a_{\nn\mu}a_{\nn\nu}^{*}a_{\mm 1\nu}^{\down}\frac{f_{\kk \nn\up}-f_{\kk \mm\up}}{E_{\kk \mm\up}-E_{\kk \nn\up}-\hbar\omega+\ii\eta}
+{a_{\mm 2\mu}^{\up}}^{*}a_{\nn\mu}a_{\nn\nu}^{*}a_{\mm 2\nu}^{\down}\frac{f_{\kk \nn\up}-f_{\kk \mm\down}}{E_{\kk \mm\down}-E_{\kk \nn\up}-\hbar\omega+\ii\eta})
\end{equation}
with $a_{\mm s\mu}^{\sigma}=\langle\kk\mu\sigma\vert\kk\mm s\rangle, s=1,2,$ and $a_{\nn\mu}=\langle\kk\mu\vert\kk\nn\rangle$ is independent of spin. Since $a_{\mm 1\nu}^{\down}\sim\xi$ we take $a_{\mm 1\mu}^{\up}=a_{\mm\mu}$ in the first term of~\eqref{chi03}. Also  $\sum_{\mu}a_{\mm\mu}^{*}a_{\nn\mu}=\delta_{\mm\thinspace\nn}$ by closure so that the first term of~\eqref{chi03} vanishes since the difference of Fermi functions is zero. Only the second term of $\chi_{\up\up\up\down}^{0}$ remains and this becomes, by use of~\eqref{perturbed state2},
\begin{equation}\label{chi04}
\chi_{\up\up\up\down}^{0}=-\xi\sum_{\kk\mu\nu}\sum_{\mm\,\nn\,\pp}\frac{\langle\kk\pp\up\vert h^{so}\vert\kk\mm\down\rangle^{*}}{E_{\kk \mm\down}-E_{\kk \pp\up}}a_{\pp\mu}^{*}a_{\nn\mu}a_{\nn\nu}^{*}
a_{\mm\nu}\frac{f_{\kk \nn\up}-f_{\kk \mm\down}}{E_{\kk \mm\down}-E_{\kk\nn\up}-\hbar\omega+\ii\eta}.
\end{equation}
Again using closure only terms with $\pp=\mm=\nn$ survive and the matrix element of $h^{so}$ becomes
\begin{equation}\label{LS}
\langle\kk\nn\up\vert\sum_{j}\mathbf{L}_{j}\cdot \mathbf{S}_{j}\vert\kk\nn\down\rangle
=\frac{1}{2}\langle\kk\nn\vert L^{-}\vert\kk\nn\rangle=0 
\end{equation} 
due to the quenching of total orbital angular momentum $\mathbf{L}=\sum_{j}\mathbf{L}_j$~\cite{CE97}. Thus, to first order in $\xi$, $\chi_{\up\up\up\down}^{0}(0,\omega)$, and similar response functions with one reversed spin, are zero. Hence we have only to solve $1-U\chi_{\down\up\up\down}^{0}=0$ to obtain $\Im(\hbar\omega)$ to $O(\xi^{2})$. Here we assume the system has spatial inversion symmetry without which the quenching of orbital angular momentum, as expressed by ~\eqref{LS}, no longer pertains~\cite{CE97}. We briefly discuss the consequences of a breakdown of inversion symmetry at the end of this section. 

On introducing the perturbed states~\eqref{perturbed state1} and~\eqref{perturbed state2} we write~\eqref{chi02} in the form
\begin{equation}\label{chitransverse0B}
\begin{split}
\chi_{\down\up\up\down}^{0}(0,\omega)=\frac{1}{N}\sum_{\kk\mm\,\nn}(\vert B_{\mm1\nn1}^{\down\up}\vert^{2}\frac{f_{\kk\nn1}-f_{\kk\mm1}}
{E_{\kk\mm1}-E_{\kk\nn1}-\hbar\omega+\ii\eta}+\vert B_{\mm1\nn2}^{\down\up}\vert^{2}\frac{f_{\kk\nn2}-f_{\kk\mm1}}
{E_{\kk\mm1}-E_{\kk\nn2}-\hbar\omega+\ii\eta}\\+\vert B_{\mm2\nn1}^{\down\up}\vert^{2}\frac{f_{\kk\nn1}-f_{\kk\mm2}}
{E_{\kk\mm2}-E_{\kk\nn1}-\hbar\omega+\ii\eta}+\vert B_{\mm2\nn2}^{\down\up}\vert^{2}\frac{f_{\kk\nn2}-f_{\kk\mm2}}
{E_{\kk\mm2}-E_{\kk\nn2}-\hbar\omega+\ii\eta}).
\end{split}
\end{equation}
Clearly $B_{\mm1\nn1}^{\down\up}$ and $B_{\mm2\nn2}^{\down\up}$ are of order $\xi$,  $B_{\mm1\nn2}^{\down\up}$ is $O(\xi^{2})$ and  $B_{\mm2\nn1}^{\down\up}$ is $O(1)$. We therefore neglect the term $\vert B_{\mm1\nn2}^{\down\up}\vert^{2}$ and, using~\eqref{perturbed state1} and~\eqref{perturbed state2}, we find
\begin{equation}\label{Bs}
 B_{\mm1\nn1}^{\down\up}=- B_{\mm2\nn2}^{\down\up}=\frac{\xi}{2}\frac{\langle\kk\mm\vert L^{-}\vert\kk\nn\rangle}{E_{\kk\nn\down}-E_{\kk\mm\up}}.
\end{equation}
The evaluation of $\vert B_{\mm2\nn1}^{\down\up}\vert^{2}$ requires more care. It appears at first sight that to obtain this to $O(\xi^{2})$ we need to include second order terms in the perturbed eigenstates given by~\eqref{perturbed state1} and~\eqref{perturbed state2}. However it turns out that these terms do not in fact contribute to $\vert B_{\mm2\nn1}^{\down\up}\vert^{2}$ to $O(\xi^{2})$ so we shall not consider them further. Then we find
\begin{equation}\label{Bdownup}
B_{\mm2\nn1}^{\down\up}=\delta_{\mm\,\nn}-\xi\frac{\langle\kk\mm\vert L^{z}\vert\kk\nn\rangle}{E_{\kk\nn}-E_{\kk\mm}}
-\frac{\xi^{2}}{4}\sum_{\pp}\frac{\langle\kk\mm\vert L^{z}\vert\kk\pp\rangle\langle\kk\pp\vert\ L^{z}\vert\kk\nn\rangle}{(E_{\kk\mm}-E_{\kk\pp})(E_{\kk\nn}-E_{\kk\pp)}}
\end{equation}
and hence to $O(\xi^{2})$
\begin{equation}\label{Bsquared}
\vert B_{\mm2\nn1}^{\down\up}\vert^{2}=\delta_{\mm\,\nn}(1-\frac{\xi^{2}}{2}\sum_{\pp}
\frac{\vert\langle\kk\mm\vert L^{z}\vert\kk\pp\rangle\vert^{2}}{(E_{\kk\mm}-E_{\kk\pp})^{2}})
+\xi^{2}\frac{\vert\langle\kk\mm\vert L^{z}\vert\kk\nn\rangle\vert^{2}}{(E_{\kk\nn}-E_{\kk\mm})^{2}}
\end{equation}
The contribution of this quantity to $\chi_{\down\up\up\down}^{0}(0,\omega)$ in~\eqref{chitransverse0B} may be written to $O(\xi^{2})$ as 
\begin{equation}\label{Bsquaredcontribution}
\frac{1}{N}\sum_{\kk\mm\,\nn}\vert B_{\mm2\nn1}^{\down\up}\vert^{2}\frac{f_{\kk\nn1}-f_{\kk\mm2}}
{E_{\kk\mm2}-E_{\kk\nn1}-b_{ex}+\ii\eta}(1-\frac{b_{ex}-\hbar\omega}{E_{\kk\mm2}-E_{\kk\nn1}-b_{ex}+\ii\eta}).
\end{equation}
This is obtained by introducing the identity $-\hbar\omega=-b_{ex}+(b_{ex}-\hbar\omega)$ in the relevant denominator in~\eqref{chitransverse0B}, and expanding to first order in $b_{ex}-\hbar\omega$ which turns out to be $O(\xi^{2})$. The remaining factors of this second term in~\eqref{Bsquaredcontribution} may then be evaluated with $\xi=0$, as at the beginning of this section, so that this term becomes $(\hbar\omega-b_{ex})/(2U^{2}\langle S^{z}\rangle)$.
By combining equations~\eqref{chitransverse0B},~\eqref{Bs} and~\eqref{Bsquaredcontribution}, and ignoring some real terms, we find that the equation $1-U\chi_{\down\up\up\down}^{0}(0,\omega)=0$ leads to the relation
\begin{equation}\label{Imomega}
\begin{split}
\Im{(\hbar\omega)}=\pi\xi^{2}/(2\langle S^z\rangle)\sum_{\kk\nn\,\mm}[(f_{\kk\nn\up}-f_{\kk\mm\down}) \vert\langle\kk\mm\vert L^{z}\vert\kk\nn\rangle\vert^{2}\delta(E_{\kk\mm\down}-E_{\kk\nn\up}-b_{ex})\\
+(1/4)(f_{\kk\nn\up}+f_{\kk\nn\down}-f_{\kk\mm\up}-f_{\kk\mm\down})\vert\langle\kk\mm\vert L^{-}\vert\kk\nn\rangle\vert^{2}\delta(E_{\kk\mm}-E_{\kk\nn}-b_{ex})]
\end{split}
\end{equation}
The Gilbert damping parameter $\alpha$ is given by $\Im(\hbar\omega)/b_{ex}$ (e.g.~\cite{EW09}) and in~\eqref{Imomega} we note that
\begin{equation}\label{delta1}
\begin{split}
(f_{\kk\nn\up}-f_{\kk\mm\down})\delta(E_{\kk\mm\down}-E_{\kk\nn\up}-b_{ex})
=[F(E_{\kk\nn\up}-\mu_{0})-F(E_{\kk\nn\up}+b_{ex}-\mu_{0})]\delta(E_{\kk\mm\down}-E_{\kk\nn\up}-b_{ex})\\
=b_{ex}\delta(E_{\kk\nn\up}-\mu_{0})\delta(E_{\kk\mm\up}-\mu_{0})
\end{split}
\end{equation}
to first order in $b_{ex}$ at temperature $T=0$. Similarly
\begin{equation}\label{delta2}
(f_{\kk\nn\sigma}-f_{\kk\mm\sigma})\delta(E_{\kk\mm}-E_{\kk\nn}-b_{ex})
=b_{ex}\delta(E_{\kk\nn\sigma}-\mu_{0})\delta(E_{\kk\mm\sigma}-\mu_{0}).
\end{equation}
Thus from~\eqref{Imomega}
\begin{equation}\label{alpha} 
\begin{split}
\alpha=\pi\xi^{2}/(2\langle S^z\rangle)\sum_{\kk\nn\,\mm} \vert\langle\kk\mm\vert L^{z}\vert\kk\nn\rangle\vert^{2}\delta(E_{\kk\nn\up}-\mu_{0})\delta(E_{\kk\mm\down}-\mu_{0})\\
+\pi\xi^{2}/(8\langle S^z\rangle)\sum_{\kk\nn\,\mm\sigma}\vert\langle\kk\mm\vert L^{-}\vert\kk\nn\rangle\vert^{2}\delta(E_{\kk\nn\sigma}-\mu_{0})\delta(E_{\kk\mm\sigma}-\mu_{0})
\end{split}
\end{equation}
correct to $O(\xi^{2})$. We note that there is no contribution from intraband terms since $\langle\kk\nn\vert\mathbf{L}\vert\kk\nn\rangle=0$. It is straight-forward to show that to $O(\xi^{2})$ this is equivalent to the expression 
\begin{equation}\label{alphaKambersky}
\alpha=\pi/(2\langle S^z\rangle)\sum_{\kk\nn\,\mm}\sum_{\sigma\sigma^{\prime}}
\vert A_{\mm\sigma,\nn\sigma^{\prime}}(\kk)\vert^{2}
\delta(E_{\kk\mm\sigma}-\mu_{0})\delta(E_{\kk\nn\sigma^{\prime}}-\mu_{0})
\end{equation}
where
\begin{equation}\label{A}
A_{\mm\sigma,\nn\sigma^{\prime}}(\kk)=\xi\langle\kk\mm\sigma\vert[S^{-},h^{so}]\vert\kk\nn\sigma^{\prime}\rangle
\end{equation}
and $S^{-}$ is the total spin operator $\sum_{j}S_{j}^{-}$ with $S_{j}^{-}=S_{j}^{x}-\ii S_{j}^{y}$. This may be written more concisely as
\begin{equation}\label{alphaKambersky1}
\alpha=\pi/(2\langle S^z\rangle)\sum_{\kk nm}\vert A_{mn}(\kk)\vert^{2}
\delta(E_{\kk m}-\mu_{0})\delta(E_{\kk n}-\mu_{0})
\end{equation}
with
\begin{equation}\label{A1}
A_{mn}(\kk)=\xi\langle\kk m\vert[S^{-},h^{so}]\vert\kk n\rangle
\end{equation}
and the understanding that the one-electron states $\kk m, \kk n$ are calculated in the absence of SOC. Equation \eqref{alphaKambersky1} is the standard form of the Kambersky formula  (~\cite{K76},~\cite{GMone09}) but in the literature SOC is invariably included in the calculation of the one-electron states. This means that the intraband terms with $m=n$ no longer vanish. They involve the square of a delta-function and this problem is always addressed by invoking the effect of impurity and/or phonon scattering to replace the delta-functions by Lorentzians of width proportional to an inverse relaxation time parameter $\tau^{-1}$. Then as one approaches a perfect crystal ($\tau\rightarrow\infty$) the intraband contribution to $\alpha$ tends to infinity. This behaviour is illustrated in many papers (~\cite{GIS07},~\cite{GMtwo09},~\cite{S12},~\cite{BCEU14}). In fig. 1 of~\cite{BCEU14} it is shown clearly that $\alpha$ remains finite if one does not include SOC in calculating the one-electron states. However the effect of not including SOC is not confined to total removal of the intraband contribution. The remaining interband contribution is increased considerably in the low scattering rate regime, by almost an order of magnitude in the case of Fe.  This makes $\alpha$ almost independent of scattering rate in Fe which may relate to its observed temperature independence~\cite{BL74}. The corresponding effect in Co is insufficient to produce the increase of $\alpha$ at low scattering rate inferred from its temperature dependence.
The non-inclusion of SOC in calculating the one-electron states used in the Kambersky formula clearly makes a major qualitative and quantitative change in the results. This occurs as soon as intraband terms become dominant in calculations where they are included. For Fe, Co and Ni this corresponds to impurity content and temperature such that the scattering rate $1/\tau$ due to defects and/or phonons is less that about $10^{14} \mathrm{ sec^{-1}}$ (~\cite{BCEU14,GIS07}). Typically these metals at room temperature find themselves well into the high scattering-rate regime where the damping rate can be reliably estimated from the Kambersky interband term, with or without SOC included in the band structure~\cite{GGMS11}. The physics at room temperature is not particularly interesting. One needs to lower the temperature into the low scattering-rate regime where intraband terms, if they exist, will dominate and lead to an anomalous $\xi^3$ dependence of the damping on spin-orbit parameter $\xi$ (~\cite{K76},~\cite{GIS08},~\cite{BCEU14}). The origin of this behaviour is explained in ~\cite{K76},~\cite{BCEU14}. It arises in the $\kk$ sum of  \eqref{alphaKambersky1} from a striplike region on the Fermi surface around a line where two different energy bands cross each other in the absence of SOC. The strip width is proportional to $\xi$ , or more precisely $\vert\xi\vert$. Since $A_{nn}(\kk)$ is of order $\xi$ the contribution of intraband terms in  \eqref{alphaKambersky1} is proportional to $\vert\xi\vert^3$. Thus the intraband terms lead to terms in $\alpha$ which diverge in the limit $\tau^{-1}\rightarrow 0$ and are non-analytic functions of $\xi$. The calculation of $\alpha$ in this section can be extended to higher powers of $\xi$ than the second. No intraband terms appear and the result is an analytic power series containing only even powers of $\xi$.

The interband term in Kambersky's formula can be given a very simple 
interpretation in terms of Fermi's "golden rule" for transition probability~\cite{K07}. This corresponds to second order perturbation theory in the spin-orbit interaction. The decay of a uniform mode ($\qq=0$) magnon into an electron-hole pair involves the transition of an electron from an occupied state to an unoccupied state of the same wave-vector. This is necessarily an interband transition and the states involved in the matrix element are unperturbed, that is calculated in the absence of SOC. A quite different approach has been adopted to try and find a physical interpretation of Kambersky's intraband term (~\cite{K07,GIS08}). This employs Kambersky's earlier "breathing Fermi surface" model (~\cite{K70, KK02}) whose range of validity is uncertain.

We now briefly discuss the consequences of a breakdown of spatial inversion symmetry so that total orbital angular momentum is not quenched. In general response functions such as  $\chi_{\up\up\up\down}^{0}(0,\omega)$ with one reversed spin are no longer zero to first order in $\xi$. Hence $S_{11}$ is not given to order $\xi^2$ just by the first two terms of \eqref{S11} but involves further terms which depend explicitly on the long-range Coulomb interaction. Consequently $\alpha$ has a similar dependence which does not emerge from the torque-correlation approach. In Appendix~\ref{Ka} it is pointed out how the direct proof of the Kambersky formula breaks down in the absence of spatial inversion symmetry.
    
\section{Experimental aspects}\label{expt}

The inclusion of intraband terms in the Kambersky formula, despite their singular nature, has gained acceptance because they appear to explain a rise in intrinsic damping parameter $\alpha$ at low temperature which is observed in some systems~\cite{BL74}. The calculated intraband contribution to $\alpha$ is proportional to the relaxation time $\tau$ and it is expected that, due to electron-phonon scattering, $\tau$ will increase as the temperature is reduced. This is in qualitative agreement with data~\cite{BL74} for Ni and hcp Co. Also a small $10\%$ increase in $\alpha$ is observed in $\mathrm{Co_2 FeAl}$ films as the temperature is decreased from 300 K to 80 K~\cite{Y14}. However in Fe the damping $\alpha$ is found to be independent of temperature down to 4 K~\cite{BL74}. Very recent measurements~\cite{M15} on FePt films, with varying antisite disorder $x$ introduced into the otherwise well-ordered structure, show that $\alpha$ increases steadily as $x$ increases from 3 to $16\%$. Hence $\alpha$ increases monotonically with scattering rate $1/\tau$ as expected from the Kambersky formula in the absence of intraband terms. Furthermore for $x=3\%$ it is found that $\alpha$ remains almost unchanged when the temperature is decreased from 200 to 20 K. Ma et al~\cite{M15} therefore conclude that there is no indication of an intraband term in $\alpha$. From the present point of view the origin of the observed low temperature increase of $\alpha$ in Co and Ni is unclear. Further experimental work to confirm the results of Bhagat and Lubitz~\cite{BL74} is desirable.

The second unusual feature of the intraband term in Kambersky's formula for $\alpha$ is its $\vert\xi\vert^3$ dependence on the SOC parameter $\xi$. This contrasts with the $\xi^2$ dependence of the interband contribution which has been observed in a number of alloys at room temperature~\cite{S07}. Recently this behaviour has been seen very precisely in $\mathrm{FePd_{1-x} Pt_x}$ alloys where $\xi$ can be varied over a wide range by varying $x$~\cite{H13}.
Unfortunately this work has not been extended to the low temperature regime where the $\vert\xi\vert^3$ dependence, if it exists, should be seen. It would be particularly interesting to see low temperature data for  $\mathrm{NiPd_{1-x}Pt_x}$ and  $\mathrm{CoPd_{1-x}Pt_x}$ since it is in Ni and Co where the intraband contribution has been invoked to explain the low temperature behaviour of $\alpha$. From the present point of view, with the intraband term absent, one would expect $\xi^2$ behaviour over the whole temperature range.  
\section{Conclusions and outlook}\label{discussion}

In this paper we analyse two methods which are used in the literature to calculate the damping in magnetization dynamics due to spin-orbit coupling. The first common approach is to employ Kambersky's~\cite{K76} formula for the Gilbert damping parameter $\alpha$ which delivers an infinite value for a pure metal if used beyond second order in the spin-orbit parameter $\xi$.
The second approach~\cite{CM15} is to calculate numerically the line-width of the ferromagnetic resonance seen in the uniform transverse spin susceptibility. This is always found to be finite, corresponding to finite $\alpha$. We resolve this apparent inconsistency between the two methods by an analytic treatment of the Costa-Muniz approach for the simplified model of a ferromagnetic metal  with d-bands only. It is shown that this method leads to the Kambersky result correct to second order in $\xi$ but Kambersky's intraband scattering term, taking the non-analytic form $\vert\xi\vert^3$, is absent. Higher order terms in the present work are analytic even powers of $\xi$. The absence of Kambersky's intraband term is the main result of this paper and it is in agreement with the conclusion that Ma et al~\cite{M15} draw from their experiments on FePt films. Further experimental work on the dependence of damping on electron scattering-rate and spin-orbit parameter in other systems is highly desirable.

A secondary conclusion is that beyond second order in $\xi$ some additional physics arises which has not been remarked on previously. This is the role of long-range Coulomb interaction which is essential for a proper treatment of the longitudinal susceptibility and charge response to which the transverse susceptibility is coupled by spin-orbit interaction. Costa and Muniz~\cite{CM15} stress this coupling but fail to introduce the long-range Coulomb interaction. Generally, however, it seems unnecessary to go beyond second order in $\xi$~\cite{S07, H13} and for most bulk systems Kambersky's formula, with electron states calculated in the absence of SOC, should be adequate. However in systems without spatial inversion symmetry, which include layered structures of practical importance, the Kambersky formulation may be inadequate even to second order in $\xi$. The long-range Coulomb interaction can now play a role.  

An important property of ferromagnetic systems without inversion symmetry is the Dzyaloshinskii-Moriya interaction (DMI) which leads to an instability of the uniform ferromagnetic state with the appearance of a spiral spin structure or a skyrmion structure. This has been studied extensively in bulk crystals like MnSi~\cite{G06} and in layered structures~\cite{vonB14}. The spiral instability appears as a singularity in the transverse susceptibility $\chi(\qq,0)$ at a value of $q$ related to the DMI parameter. The method of this paper has been used to obtain a novel closed form expression for this parameter which will be reported elsewhere.  
   
In this paper we have analysed in some detail the transverse spin susceptibility $\chi_{\down\up\up\down}$ but combinations of some of the 15 other response functions merit further study. Mixed charge-spin response arising from spin-orbit coupling is of particular interest for its relation to phenomena like the spin-Hall effect. 

\appendix
\section{A direct derivation of the Kambersky formula}\label{Ka}
In this appendix we give a rather general derivation of the Kambersky formula for the Gilbert damping parameter $\alpha$ with an emphasis on its restriction to second order in the spin-orbit interaction parameter $\xi$.

We consider a general ferromagnetic material described by the many-body Hamiltonian
\begin{equation}\label{ham}
H=H_1+H_{int}+H_{ext}
\end{equation}
where $H_1$ is a one-electron Hamiltonian of the form
\begin{equation}\label{H1}
H_1=H_{k}+H^{so}+V.
\end{equation}
Here $H_{k}$ is the total kinetic energy, $H^{so}=\xi h^{so}$ is the spin-orbit interaction, $V$ is a potential term, $H_{int}$ is the Coulomb interaction between electrons and $H_{ext}$ is due to an external magnetic field $B_{ex}$ in the $z$ direction. Thus $H_{ext}=-S^{z}b_{ex}$ where $ b_{ex}=2\mu_{B}B_{ex}$, as in~\eqref{Vsigmaeff}, and $S^{z}$ is the $z$ component of total spin. Both $H^{so}$ and $V$ can contain disorder although in this paper we consider a perfect crystal. Following the general method of Edwards and Fisher~\cite{EF71} we use equations of motion to find that the dynamical transverse susceptibility $\chi(\omega)=\chi_{-+}(0,\omega)$ satisfies~\cite{EW09}
\begin{equation}\label{EWeqn}
\chi(\omega)=-\frac{2\langle S^{z}\rangle}{\hbar\omega-b_{ex}}+\frac{\xi^{2}}{(\hbar\omega-b_{ex})^{2}}
(\chi_{F}(\omega)-\xi^{-1}\langle[F^{-},S^{+}]\rangle)
\end{equation}
where
\begin{equation}\label{chiF}
\chi_{F}(\omega)=\int\langle\langle F^{-}(t),F^{+}\rangle\rangle e^{-\ii\omega t}dt
\end{equation}
with $F^{-}=[S^{-},h^{so}]$. This follows since $S^{-}$ commutes with other terms in $H_1$ and with $H_{int}$. For small $\omega$, $\chi$ is dominated by the spin wave pole at $\hbar\omega=b_{ext}+\hbar\delta\omega$ where $\delta\omega\sim\xi^{2}$, so that
\begin{equation}\label{chipole}
\chi(\omega)=-\frac{2\langle S^{z}\rangle}{\hbar(\omega-\delta\omega)-b_{ex}}.
\end{equation}
Following~\cite{EW09} we compare~\eqref{EWeqn} and~\eqref{chipole} in the limit $\hbar\delta\omega\ll\hbar\omega-b_{ex}$ to obtain
\begin{equation}\label{deltaomega}
-2\langle S^{z}\rangle\hbar\delta\omega=\xi^{2}(\chi_{F}(\omega)-\xi^{-1}\langle[F^{-},S^{+}]\rangle)
=\xi^{2}[\lim_{\hbar\omega\to b_{ex}}\chi_{F}^{\xi=0}(\omega)-\lim_{\xi\to 0}(\frac{1}{\xi}\langle[F^{-},S^{+}]\rangle)]
\end{equation}
correct to order $\xi^{2}$. It is important to note that the limit $\xi\rightarrow 0$ within the bracket must be taken before putting $\hbar\omega=b_{ex}$. If we put $\hbar\omega=b_{ex}$ first it is clear from~\eqref{EWeqn} that the quantity in brackets would vanish, giving the incorrect result $\delta\omega=0$. Furthermore it may be shown [M. Cinal, private communication] that the second term in the bracket is real. Hence
\begin{equation}\label{Imomega2}
\Im(\hbar\omega)=-\frac{\xi^{2}}{2\langle S^{z}\rangle}\lim_{\hbar\omega\to b_{ex}}\Im[\chi_{F}^{\xi=0}(\omega)].
\end{equation}
Kambersky~\cite{K76} derived this result, using the approach of Mori and Kawasaki (~\cite{M65}~\cite{MK62}), without noting its restricted validity to second order in $\xi$. This restriction is crucial since, as discussed in the main paper, it avoids the appearance of singular intraband terms.  Oshikawa and Affleck emphasise strongly a similar restriction in their related work on electron spin resonance (Appendix of ~\cite{OA02}).

Equation~\eqref{Imomega2} is an exact result even in the presence of disorder in the potential and spin-orbit terms of the Hamiltonian. In the following we assume translational symmetry.

To obtain the expression~\eqref{alpha} for $\alpha=\Im(\hbar\omega)/b_{ex}$, which is equivalent to Kambersky's result~\eqref{alphaKambersky}, it is necessary to evaluate the response function $\chi_{F}^{\xi=0}(\omega)$ in tight-binding-RPA. Using~\eqref{Hso}we find
\begin{equation}\label{Fminus}
F^{-}=\sum_{\kk\mu\nu}[L_{\mu\nu}^{z}c_{\kk\mu\down}^{\dagger}c_{\kk\nu\up}
+(1/2)L_{\mu\nu}^{-}(c_{\kk\mu\down}^{\dagger}c_{\kk\nu\down}-c_{\kk\mu\up}^{\dagger}c_{\kk\nu\up})].
\end{equation}
Hence
\begin{equation}\label{chiF}
\chi_{F}^{\xi=0}=\sum_{\mu\nu}\sum_{\alpha\beta}[L_{\mu\nu}^{z}L_{\beta\alpha}^{z}
G_{\mu\down\nu\up,\beta\up\alpha\down}+(1/4)L_{\mu\nu}^{-}L_{\beta\alpha}^{+}
(G_{\mu\down\nu\down,\beta\down\alpha\down}+G_{\mu\up\nu\up,\beta\up\alpha\up}
-G_{\mu\down\nu\down,\beta\up\alpha\up}-G_{\mu\up\nu\up,\beta\down\alpha\down})]
\end{equation}
where
\begin{equation}\label{G}
G_{\mu\sigma\nu\sigma^{\prime},\beta\tau\alpha\tau^{\prime}}
=\langle\langle\sum_{\kk}c_{\kk\mu\sigma}^{\dagger}c_{\kk\nu\sigma^{\prime}};
\sum_{\uu}c_{\uu\beta\tau}^{\dagger}c_{\uu\alpha\tau^{\prime}}\rangle\rangle_{\omega}.
\end{equation}
The Green function $G$ is to be calculated in the absence of SOC ($\xi=0$). Within RPA it satisfies an equation of the form
\begin{equation}\label{GRPA}
G_{\mu\sigma\nu\sigma^{\prime},\beta\tau\alpha\tau^{\prime}}=G_{\mu\sigma\nu\sigma^{\prime},\beta\tau\alpha\tau^{\prime}}^{0}
-\sum_{\mu_{1}\sigma_{1}\nu_{1}\sigma_{1}^{\prime}}\,\sum_{\mu_{2}\sigma_{2}\nu_{2}\sigma_{2}^{\prime}}
G_{\mu\sigma\nu\sigma^{\prime},\mu_{1}\sigma_{1}\nu_{1}\sigma_{1}^{\prime}}^{0}
V_{\mu_{1}\sigma_{1}\nu_{1}\sigma_{1}^{\prime},\mu_{2}\sigma_{2}\nu_{2}\sigma_{2}^{\prime}}
G_{\mu_{2}\sigma_{2}\nu_{2}\sigma_{2}^{\prime},\beta\tau\alpha\tau^{\prime}}
\end{equation}
where $G^{0}$ is the non-interacting (Hartree-Fock) Green function and
\begin{equation}\label{VRPA}
V_{\mu_{1}\sigma_{1}\nu_{1}\sigma_{1}^{\prime},\mu_{2}\sigma_{2}\nu_{2}\sigma_{2}^{\prime}}
=V_{\sigma_{1}\sigma_{1}^{\prime}\sigma_{2}\sigma_{2}^{\prime}}(\qq)\delta_{\mu_{1}\nu_{1}}\delta_{\mu_{2}\nu_{2}}
\end{equation}
with $V(\qq)$ given by~\eqref{Vq} and~\eqref{K}.
Hence
\begin{equation}\label{RPA}
G_{\mu\sigma\nu\sigma^{\prime},\beta\tau\alpha\tau^{\prime}}=G_{\mu\sigma\nu\sigma^{\prime},\beta\tau\alpha\tau^{\prime}}^{0}
-\sum_{\mu_{1}\sigma_{1}\sigma_{1}^{\prime}}\,\sum_{\mu_{2}\sigma_{2}\sigma_{2}^{\prime}}
G_{\mu\sigma\nu\sigma^{\prime},\mu_{1}\sigma_{1}\mu_{1}\sigma_{1}^{\prime}}^{0}
V_{\sigma_{1}\sigma_{1}^{\prime}\sigma_{2}\sigma_{2}^{\prime}}
G_{\mu_{2}\sigma_{2}\mu_{2}\sigma_{2}^{\prime},\beta\tau\alpha\tau^{\prime}}.
\end{equation}
The form of the interaction $V$ given in~\eqref{VRPA} is justified by the connection between~\eqref{RPA} and~\eqref{chiq}, with $q=0$. To see this connection we note that $\chi_{\sigma\sigma^{\prime}\tau\tau^{\prime}}
=\sum_{\mu\nu}G_{\mu\sigma\mu\sigma^{\prime},\nu\tau\nu\tau^{\prime}}$ and that~\eqref{RPA} then leads to~\eqref{chiq} which is equivalent to RPA. On substituting~\eqref{RPA} into~\eqref{chiF} we see that the contributions from the second term of~\eqref{RPA} contain factors of the form
\begin{equation}\label{vanishing factors}
\sum_{\mu\nu\mu_{1}}L_{\mu\nu}^{z}G_{\mu\down\nu\up,\mu_{1}\up\mu_{1}\down}^{0},\quad
\sum_{\mu\nu\mu_{1}}L_{\mu\nu}^{-}G_{\mu\sigma\nu\sigma,\mu_{1}\sigma_{1}\mu_{1}\sigma_{1}}^{0}.
\end{equation}
We now show that such factors vanish owing to quenching of orbital angular momentum in the system without SOC ($\xi=0$). Hence the Green functions $G$ in~\eqref{chiF} can be replaced by the non-interacting ones $G^{0}$. The non-interacting Green functions $G^{0}$ are of a similar form to $\chi^{0}$ in~\eqref{chi0explicit} and for $\xi=0$ may be expressed in terms of the quantities $a_{\nn\mu}=\langle\kk\mu\vert\kk\nn\rangle$ where $\vert\kk\nn\rangle$ is a one-electron eigenstate as introduced in section~\ref{linewidth}. Hence we find, in the same way that~\eqref{transversechi0} emerged,
\begin{equation}\label{factor}
\sum_{\mu\nu\mu_{1}}L_{\mu\nu}^{z}G_{\mu\down\nu\up,\mu_{1}\up\mu_{1}\down}^{0}
=\sum_{\mu\nu}\sum_{\kk\nn}L_{\mu\nu}^{z}a_{\nn\mu}^{*}a_{\nn\nu}
\frac{f_{\kk\nn\up}-f_{\kk\nn\down}}{\Delta+b_{ex}-\hbar\omega+\ii\eta}.
\end{equation}
Also by closure
\begin{equation}\label{factor0}
\sum_{\mu\nu}L_{\mu\nu}^{z}a_{\nn\mu}^{*}a_{\nn\nu}=\langle\kk\nn\vert L^{z}\vert\kk\nn\rangle=0,
\end{equation}
the last step following from quenching of total orbital angular momentum. The proof that the second expression in~\eqref{vanishing factors} vanishes is very similar.

Hence we can insert the non-interacting Green functions $G^{0}$ in~\eqref{chiF} and straight-forward algebra, with use of~\eqref{Imomega2}, leads to~\eqref{Imomega}. At the end of section~\ref{linewidth} this is shown to be equivalent to the Kambersky formula for $\alpha$. We emphasize again that the present proof is valid only to order $\xi^{2}$ so that the one-electron states used to evaluate the formula should be calculated in the absence of SOC.

This proof relies on the quenching of orbital angular momentum which does not occur in the absence of spatial inversion symmetry. When this symmetry is broken it is not difficult to see that the second term of ~\eqref{RPA} gives a contribution to the first term on the right of ~\eqref{chiF} which contains the $\qq=0$ spin-wave pole and diverges as  $\hbar\omega\rightarrow b_{ex}$. Hence the proof of the torque-correlation formula ~\eqref{Imomega2}  collapses. The method of section~\ref{linewidth} must be used as discussed at the end of that section.
\section{Elements of S}\label{Selements}
The element $S_{11}$ of the matrix $S$ is given in~\eqref{S11}. The remaining elements are given below.
\begin{equation}\label{S12}
\begin{split}
S_{12}=-U\chi_{\down\up\down\up}^{0}
+(U/\Lambda)[(X+\chi_{\down\down\down\up}^{0})\chi_{\up\up\down\up}^{0}\chi_{\down\up\up\up}^{0}
-(Y+\chi_{\up\up\down\down}^{0})\chi_{\down\down\down\up}^{0}\chi_{\down \up \up\up}^{0}\\-(Y+\chi_{\down\down\up\up}^{0})\chi_{\up\up\down\up}^{0}\chi_{\down\up\down\down}^{0}
+(X+\chi_{\up\up\up\up}^{0})\chi_{\down\down\down\up}^{0}\chi_{\down\up\down\down}^{0}]
\end{split}
\end{equation}
\begin{equation}\label{S21}
\begin{split}
S_{21}=-U\chi_{\up\down\up\down}^{0}
+(U/\Lambda)[(X+\chi_{\down\down\down\down}^{0})\chi_{\up\up\up\down}^{0}\chi_{\up\down\up\up}^{0}
-(Y+\chi_{\up\up\down\down}^{0})\chi_{\down\down\up\down}^{0}\chi_{\up \down \up\up}^{0}\\-(Y+\chi_{\down\down\up\up}^{0})\chi_{\up\up\up\down}^{0}\chi_{\up\down\down\down}^{0}
+(X+\chi_{\up\up\up\up}^{0})\chi_{\down\down\up\down}^{0}\chi_{\up\down\down\down}^{0}]
\end{split}
\end{equation}
\begin{equation}\label{S21}
\begin{split}
S_{22}=1-U\chi_{\up\down\down\up}^{0}
+(U/\Lambda)[(X+\chi_{\down\down\down\down}^{0})\chi_{\up\up\down\up}^{0}\chi_{\up\down\up\up}^{0}
-(Y+\chi_{\up\up\down\down}^{0})\chi_{\down\down\down\up}^{0}\chi_{\up \down \up\up}^{0}\\-(Y+\chi_{\down\down\up\up}^{0})\chi_{\up\up\down\up}^{0}\chi_{\up\down\down\down}^{0}
+(X+\chi_{\up\up\up\up}^{0})\chi_{\down\down\down\up}^{0}\chi_{\up\down\down\down}^{0}]
\end{split}
\end{equation}

\section*{Acknowledgement}
My recent interest in Gilbert damping arose through collaboration with O. Wessely, E. Barati, M. Cinal and A. Umerski. I am grateful to them for stimulating discussion and correspondence. The specific work reported here arose directly  from discussion with R.B.Muniz and I am particularly grateful to him and his colleague A. T. Costa for this stimulation. 

\section*{References}
%\bibliographystyle{iopart-num}
%\bibliography{ref}

\end{document}